\documentclass{aa}
\usepackage{graphics}
\usepackage{times}

\def\wig#1{\mathrel{\hbox{\hbox to 0pt{%
\lower.5ex\hbox{$\sim$}\hss}\raise.4ex\hbox{$#1$}}}}
\def\msol{\mbox{M}_\odot}

\def\kms{\,{\rm km.s}^{-1}}
\def\mssurf{\mbox{M}_\odot\, {\rm pc}^{-2}}
\def\msvol{\mbox{M}_\odot\, {\rm pc}^{-3}}
\def\vrot{v_{\rm rot}}
\def\beq{\begin{equation}}
\def\eeq{\end{equation}}

\begin{document}

\def\aj{AJ}                  
\def\araa{ARA\&A}             
\def\apj{ApJ}                 
\def\apjl{ApJ}                
\def\apjs{ApJS}               
\def\ao{Appl.Optics}          
\def\apss{Ap\&SS}             
\def\aap{A\&A}                
\def\aanda{A\&A}                
\def\aandal{A\&AL}                
\def\aapr{A\&A~Rev.}          
\def\aaps{A\&AS}             
\def\aandas{A\&AS}             
\def\azh{AZh}                 
\def\baas{BAAS}               
\def\jrasc{JRASC}             
\def\memras{MmRAS}            
\def\mnras{MNRAS}             
\def\pra{Phys.Rev.A}          
\def\prb{Phys.Rev.B}          
\def\prc{Phys.Rev.C}          
\def\prd{Phys.Rev.D}          
\def\prl{Phys.Rev.Lett}       
\def\pasp{PASP}               
\def\pasj{PASJ}               
\def\qjras{QJRAS}             
\def\skytel{S\&T}             
\def\solphys{Solar~Phys.}     
\def\sovast{Soviet~Ast.}      
\def\ssr{Space~Sci.Rev.}      
\def\zap{ZAp}                 

\thesaurus{}

\title{Towards a consistent model of the Galaxy : II. Derivation of the
model}

\author{{\sc D. M\'era$^{1,3}$, G. Chabrier$^1$ and R. Schaeffer$^2$}}

\institute{$^1$ C.R.A.L. (UMR CNRS 5574),
Ecole Normale Sup\'erieure, 69364 Lyon Cedex 07, France,\\
$^2$Service de Physique Th\'eorique,
CEA Saclay, 91191 Gif-sur-Yvette, France\\
$^3$Physics department, Whichita State University, 1845 Fairmount, Wichita KS 67260, USA}

\date{Received date ; accepted date}

\maketitle

\markboth{}{M\'era, Chabrier and Schaeffer: Consistent model of the
Galaxy II}

\begin{abstract} 

We use the calculations derived in a previous paper (M\'era, Chabrier
and Schaeffer, 1997)\nocite{Meraetal97a}, based on observational
constraints arising from star counts, microlensing experiments and
kinematic properties, to determine the amount of dark matter under
the form of stellar and sub-stellar objects in the different parts of
the Galaxy.  This yields the derivation of different mass-models for
the Galaxy. In the light of {\it all} the afore-mentioned constraints,
we discuss two models that correspond to different conclusions about
the nature and the location of the Galactic dark matter. In the first
model there is a small amount of dark matter in the disk, and a large
fraction of the dark matter in the halo is still undetected and likely
to be non-baryonic.  The second, less conventional model is consistent
with entirely, or at least predominantly baryonic dark matter, under the
form of brown dwarfs in the disk and white dwarfs in the dark halo. We
derive observational predictions for these two models which should be
verifiable by near future infrared and microlensing observations.

\bigskip


\end{abstract}

\section{Introduction}

One of the outstanding problems in astrophysics is the nature of dark
matter in the Universe. Over a wide range of scales the dynamics of
gravitationally bound systems reveals the presence of a large, dominant
amount of dark material.  There are in fact two dark matter problems.
The ``large-scale'' dark matter that is the main contributor to $\Omega$
is essentially non-baryonic, if $\Omega > 0.1$. And the ``galactic''
dark matter, whose presence is inferred from the small-scale (velocity
dispersion) and large-scale (circular rotation velocity) kinematic
properties in galaxies. On the other hand primordial nucleosynthesis, in
particular the concordance of the predictions of $^4$He, $^3$He, $^2$H and
$^7$Li abundances bounds $\Omega_B$, with a lower limit $\Omega_B\sim
0.01\,h^{-2}$ ($h$ is the Hubble constant in unit of 100 $\kms$
Mpc$^{-1}$) which exceeds the density of visible baryons (see e.g. Carr,
1994; Copi, Schramm and Turner, 1995).\nocite{Carr94,Copietal95}
Coincidentally, galactic halos span the range where dark matter could
be entirely baryonic, providing a halo radius $R_h\wig < 50\,h^{-1}$ kpc.

Within the past few years, microlensing observations in the central parts
and in the halo of our Galaxy shed new light on the galactic dark matter
problem by revealing the presence of dark objects in the Galaxy and
thus of ``some'' baryonic dark matter in the Milky Way. In a previous
paper (M\'era, Chabrier \& Schaeffer, 1997; paper I) we have examined
the different observational constraints arising from star counts at
faint magnitudes, kinematic properties and microlensing experiments to
derive mass functions in the Galactic disk and halo and to determine as
accurately as possible the stellar and substellar contributions to the
Galactic mass budget.

In this paper, we examine different mass-models for the Galaxy compatible
with these determinations and thus with {\it all} presently available
observational constraints. In \S2, we first discuss the nature of the
possible stellar content of the halo. In \S3, we examine what we call
the conventional scenario, with a standard massive halo and a substantial
fraction of unidentified, most likely {\it non-baryonic} matter. In \S4,
we devote some attention to the maximal disk scenario, whose interest
has been renewed in the last years.  Section 5 is devoted to the
derivation of a less conventional model, which determines the limits of
an {\it entirely-baryonic} dark matter model for the Galaxy. Summary and
conclusions are outlined in \S6.  We propose observational predictions
characteristic of these models which should be in reach of near future
infrared and microlensing projects.

\section{Which objects in the dark halo ?}

The first analysis of the Hubble Space Telescope (HST) deep star counts
showed that M-dwarfs provide at most $\sim 10-20\%$ of the halo dynamical
mass (Bahcall et al., 1994; Santiago, Gilmore \& Elson, 1996). A more
detailed analysis, consistent with the observed number of high-velocity
subdwarfs in the solar neighborhood (Dahn et al, 1995) places much more
stringent limits with a main sequence star contribution to the halo mass
smaller than 1\% (M\'era, Chabrier \& Schaeffer, 1996; Graff \& Freese,
1996; Chabrier \& M\'era, 1997a; M\'era et al., 1997, paper I).  Therefore
M-dwarfs are definitely excluded as a significant halo population and
can be responsible for less than 0.1 of the 6 to 8 events\footnote{It
has been suggested by Gould et al. (1997)\nocite{GouldBahcallFlynn97}
that one of these events might be due to a {\it disk} M-dwarf.} observed
towards the LMC (see paper I).  This raises a severe problem for the
observed microlensing events towards the LMC. The nature of the objects
responsible for these events is still under debate, and represents a
central issue for any model of the Galaxy.

The optical depth inferred from the LMC observations is $\tau_{\rm
obs}=2.2\,\pm\,1\times 10^{-7}$, and the average mass for the lens
is $\langle \sqrt m\rangle^2 \approx 0.5\,\msol$ (Alcock et al.,
1996b)\nocite{Alcocketal96b}. This mass is determined with a standard,
spherically symmetric, halo model (see paper I \S4.3). The velocity
dispersion decreases with oblateness, leading to a lower average
mass. Within the range of flattened Evans models (Evans, 1993; 1994),
the characteristic mass of lenses remains larger than 0.1 $\msol$ (see
Paper I \S4.3). Only with a very flattened halo (axis ratio of the order
of 1/10), the velocity dispersion is low enough for the average mass to
lie under the hydrogen burning limit. Moreover, such a disk-like halo
must have a relatively small asymmetric drift, which yields an even
lower $\langle 1/v_\bot\rangle^{-1}$ and an average lens mass $\langle
\sqrt{m}\rangle^2\approx 0.08\msol$. But the spheroid MF normalization at
0.1 $\msol$ (Eq. (16) of paper I) is reasonably well constrained. Chabrier
and M\'era (1997a)\nocite{ChabrierMera97a} have shown that the local
spheroid subdwarfs are not related to the dark matter halo, yielding
an even more severe constraint on the dark halo MF, whose maximum
normalization is $\sim 1\%$ of the spheroid one at 0.1 $\msol$. Moreover,
the extrapolation of the spheroid MF into the brown dwarf domain down to
a minimum mass corresponding to an average mass of 0.08 $\msol$ yields
a mass density of less than 2\% of the dark halo density. Clearly,
the substellar part of that MF would have to be a Dirac delta-function
at 0.08 $\msol$ for the whole observed LMC optical depth to be due to
brown dwarfs and these brown dwarfs should be visible in the Hubble Deep
Field. In any case a highy flattened halo raises severe problems. Features
like the Magellanic Stream, for example, can not be accounted for in
such a model. Reasonable values for the halo oblateness are more in the
range $q\sim 0.6$ (e.g. Sackett et al., 1994),\nocite{Sackettetal94}
so that brown dwarfs seem to be definitely excluded as halo dark matter
candidates, unless they are strongly inhomogeneously distributed in dark
clusters (Kerins, 1997).\nocite{Kerins97}

\medskip

The expected number of microlensing events in the LMC itself is:

\beq N_{\sc LMC}\approx
\frac{\sigma}{25\,\kms}\,\sqrt{\frac{h}{250\,pc}}\,
\frac{\Sigma}{360\,\mssurf}\label{Nlmc}\eeq

\noindent where $\sigma, \Sigma$ and $h$ denote the velocity dispersion,
the surface density and the depth of the LMC \footnote{Note that these variables
are not independent from each other.}.  Reasonable values for these
parameters (Gould, 1995)\nocite{Gould95} yield $N_{\sc LMC}\sim 1$,
which is consistent with the analysis of the binary event in the recent
MACHO results (Bennett et al., 1997)\nocite{Bennettetal97}.

The only remaining (though hardly satisfying) solution to explain the
observed events toward the LMC is massive stellar remnants, white
dwarfs (WD) or neutron stars (NS), as first suggested by the MACHO
collaboration. Halo NS could be disk NS whose high-proper motion exceeds
the disk escape velocity. However, the total number of NS in the Galatic
disk amounts to $\sim 10^9$, with a NS mass strongly peaked around
$\sim 1.4\,\msol$. This yields a negligible ($<<1\%$) contribution to
the observed optical depth.  Moreover, NS are the end product of type
II supernovae, whose expected number in the Galactic halo is severely
constrained by the observed metallicity.  In any case NS's and stellar
black holes are likely to contribute considerably less than WDs to the
dark mass fraction.  If the ratio of WDs to low-mass stars in the halo
is assumed to be the same as the one observed in the disk, i.e. $\sim
1/100$ ($N_{WD_{disk}}\sim 3\times 10^{-3}\,\mbox{pc}^{-3}$, Liebert,
Dahn \& Monet, 1988) and if we assume the same amount of mass in NS -
a very upper limit -, we get a {\it maximum contribution} of
baryonic stellar remnants to the Galactic dynamic mass of $\sim 5\%$.
This corresponds to less than 1 event towards the LMC. Clearly, for WDs
to be responsible for the dark events towards the LMC, one must advocate
very particular conditions in the primordial halo.

Recently, Chabrier, Segretain \& M\'era
(1996)\nocite{ChabrierSegretainMera96} and Adams \& Laughlin
(1996)\nocite{AdamsLaughlin96} have examined in detail the possibility
for WDs to provide a significant halo population. These authors have
shown that it is indeed possible for these objects to be responsible
for the observed MACHO events, thus providing $\sim 40\%$ of the
missing mass, under two {\it necessary conditions}, namely i) a halo age
substantially older than the disk and a {\it very peculiar} halo initial
mass function (IMF) strongly peaked (or cut) around $\sim 1.5-2\,\msol$,
completely different from the one determined in globular clusters and
in the spheroid (Chabrier \& M\'era, 1997a)\nocite{ChabrierMera97a}
or in the Galactic disk (M\'era, Chabrier \& Baraffe, 1996; M\'era
et al., 1997, paper I).\nocite{MeraChabrierBaraffe96} This scenario
also implies that a substantial fraction of the helium produced
in the WD progenitors has been blown away from the Galaxy along
its evolution. This is supported by the recent suggestion of the
presence of hot, metal-rich gas in the Local Group (Suto et al., 1996;
Fields et al., 1997)\nocite{Sutoetal96,Fieldsetal97}, as observed in
other groups of galaxies, although this scenario has been questioned
recently (Gibson \& Mould, 1997).\nocite{GibsonMould97} Although the WD
hypothesis seems very speculative and raises important questions about
the nucleosynthesis of primordial intermediate-mass stars (Gibson \&
Mould, 1997)\nocite{GibsonMould97} and requires unambigous confirmation
of microlensing experiments in the Galactic halo, it cannot be excluded
a priori. As shown by Chabrier et al. (1996), it is consistent with
constraints arising from the expected radiation signature of the
progenitors at large redshift (Charlot \& Silk, 1995) and from the
observed (and unobserved !) density of high velocity white dwarfs in
the solar neighborhood (Liebert et al., 1988).\nocite{Liebertetal88}
Calculations of the expected discovery functions and star counts at faint
magnitudes for various halo white dwarf luminosity functions are under
way (Chabrier and M\'era, 1997b),\nocite{ChabrierMera97b} that will
be confronted to near-future high-proper-motion surveys. This will soon
confirm or rule out the hypothesis of a significant halo WD population.

Interestingly enough, the WD scenario also bears important
consequences on our understanding of the formation of Type Ia
supernovae. If SNI form from accreting WDs, a high-density WD
halo would imply 100 times more supernovae in the halo, with
$M_{WD}\approx 0.40\times 10^{12}\,\msol$ within 50 kpc, than in
the disk, where $M_{WD}\approx 0.2\times 10^{10}\,\msol$. Such a
high rate of SNI in the halo of our Galaxy and neighbor galaxies is
clearly excluded by the observations. In case WD's do indeed represent
a significant halo population, this may imply that SNIa do not form
from accreting WDs, as suggested by several authors (Benz et al. 1990;
Mochkovitch \& Livio, 1989, 1990; Nomoto \& Iben, 1985; Saio \& Nomoto,
1985).\nocite{Benzetal90,Mochetal89,Mochetal90,NomotoIben85,SaioNomoto85}
The final outcome of merging WDs might rather be a single, massive white
dwarf (Benz et al. 1990; Segretain, Chabrier \& Mochkovitch, 1997).
\nocite{Benzetal90,SegretainCM97} Another possibility, in case direct
observations confirm the WD hypothesis, could be that their mass function
is sufficiently narrow to prevent any instability due to accretion,
or that they are old enough so that mass exchanges in binary systems no
longer occur.

\medskip

Although other possibilities may be considered for the microlensing
objects, such as, for instance, new (unknown) variable stars, hypothetical
novae eruptions (Della Valle and Livio, 1996)\nocite{DellaValleLivio96} or
primordial black holes, white dwarfs remain the least unlikely solution.
The models for the Galaxy can thus be split into two types: the first
one postulates that the dark halo is made of matter of unknown nature
and does not necessarily require the halo WD scenario (\S3), the second
type considers the possibility for WD-like objects to explain all,
or at least most of the halo dark matter (\S5).

\section{Standard massive halo model}

The determination of the different contributions to the dark matter
density in the solar neighborhood can be understood by the following
simplified analytical approach.

For a rotation velocity independent of the galactocentric distance,
the Poisson equation implies a density distribution for the dark halo,
assuming it is the only contributor to the dynamics :

\beq\rho_h(r)=\frac{\vrot^2}{4\pi G} \frac{1}{r^2}\label{rhohmax}\eeq

\noindent where $\vrot \approx 220\,\kms$ is the asymptotic rotation
velocity (see paper I). The {\it maximal halo} solar density corresponds
to:

\beq\rho_{h \odot}^{(max)} = \frac{\vrot^2}{4\pi G} \frac{1}{R_\odot^2}
\approx 0.012 \, \msvol\label{rhohsolmax}\eeq

\noindent where $R_\odot=8.5$ kpc is the galactocentric position of
the Sun.  The disk, however, contributes to the rotation velocity in
the solar neighbourhood $v_\odot$. This implies a slightly reduced
local halo density as compared to the maximal halo (\ref{rhohsolmax}).
At large distances, the disk exponential density-profile yields a much
smaller correction. Since there is no hint for a significant drop in
the observed rotation curve up to at least 70 kpc, this compensation is
usually taken into account by introducing a screened halo density-profile:

\beq \rho_h (r) =  \ \frac{\vrot^2}{4\pi G }\frac{1}{R_c^2 +
r^2}\label{rhoh}\eeq

\noindent which reduces the halo density in the solar neighborhood to
(for $R_c=5$ kpc):

\beq\rho_{h \odot}^{(max)} = \frac{\vrot^2}{4\pi G} \frac{1}{R_c^2 +
R_\odot^2} = 9.2\times 10^{-3} \msvol\label{rhohsol}\eeq

The rotation velocity $v_\odot$ is thus related to the halo local volume
density $\rho_{h_\odot}$ and to the disk $\propto \exp(-r/R_d)$ (where
$R_d$ is the disk scale length) local surface density $\Sigma_\odot$,
approximately by (see e.g. Binney \& Tremaine, 1987):

\beq\frac{v_\odot^2}{2\pi G} = 2 f(R_c/R_\odot)R_\odot^2\rho_{h_\odot}
+ \frac{v_{\rm disk}^2}{2\pi G} + \frac{M_{\rm bulge}}{2\pi R_\odot}\,
\label{relvrhosig}\eeq

\noindent where 
\bigskip

$f(x)=[1-x\,{\rm Arctan}(1/x)] [1+x^2]$

$v_{\rm disk}^2/2\pi G=2R_d y^2 \left[
I_0(y)K_0(y)-I_1(y)K_1(y)\right]\Sigma_\odot e^{R_\odot/R_d}$

$y=R_\odot/2R_d,$
\bigskip

\noindent $I_n$ and $K_n$ are Bessel functions,
and $M_{\rm bulge}$ is the mass of the central bulge.

On the other hand, the contributions of the disk and the halo to the
vertical acceleration are given schematically by (assuming $z$ to be small
enough for $A^2-B^2\approx 0$, where $A$ and $B$ are the Oort constants):

\beq \frac{K(z)}{2\pi G}=\frac{1}{2\pi G}{\partial \phi\over
\partial z} \approx \Sigma_\odot + 2 z\, \rho_{h_\odot}
\label{Kofz}\eeq

The standard halo model corresponds to the self-consistent solution of
equations (\ref{relvrhosig}) and (\ref{Kofz}) which determines $R_c$
and $\Sigma_\odot$ from the measured $\vrot$ and $ \frac{K(z)}{2\pi
G}$. For $v_{rot} = 220\;\kms$,  $\frac{K(z)}{2\pi G}  = 71 \,\mssurf$
at $z = 1.1$ kpc, $R_\odot = 8.5$ kpc and $R_d = 3.5$ kpc, we get $R_c
= 4.1$ kpc and $\Sigma_\odot = 49\, \mssurf$, as Kuijken and Gilmore
(1991).\nocite{KuijkenGilmore89b} The exact determination of the
respective contributions to the gravitational acceleration (\ref{Kofz})
due to the disk and halo is in fact more complicated and requires detailed
(model-dependent) numerical computations (see e.g. Bienaym\'e, Robin \&
Cr\'ez\'e, 1987),\nocite{BienaymeRobinCreze87} but the present approach
yields reasonnably accurate values for the dynamical constraints to be
fulfilled by different disk/halo models. It is clear that {\it all} the
values of the disk surface density derived from the measurement of the
vertical acceleration in the literature were done within the standard
halo model, and thus include implicitly this assumption. Any Galactic
mass-model using a different (non maximal) halo will thus imply values for
the observed surface density larger than the ones currently given in the
litterature (since the halo contribution will be smaller than the present
standard one). This important issue will be examined in \S 4 and 5.

We argued in Paper I that the most recent determination of the surface
density by Flynn and Fuchs (1994), $\Sigma_\odot = 52 \pm 13 \,\mssurf$,
which we use in the present calculations, is consistent at the 1$\sigma$
level with all previous determinations and is nearly identical to the
uncertainty-weighted average $\Sigma_\odot = 51 \pm 6 \,\mssurf$ of these
different but consistent measurements. The density under the form of {\it observed}
objects, determined in paper I (Eq. 15), $\Sigma_\odot = 43 \pm 5
\,\mssurf$, is compatible at the 1$\sigma$ level with the
afore-mentioned dynamical determination, leaving limited possibility
for a brown dwarf component $\Sigma_{\rm BD} = 9 \pm 14 \,\mssurf$. If,
however, the 1$\sigma$ upper limit is retained, then the disk may
contain a brown dwarf density as high as $\Sigma_{\rm BD_\odot} \simeq
22 \,\mssurf $, {\it comparable to the main sequence star contribution},
still compatible with the microlensing observations towards the bulge,
as shown in section 4.1.2 of paper I.

The final parameters for the mass distribution in the Galaxy in the
conventional model are thus:

${\bullet}$ an exponential disk with a scale length $R_d = 3.5$ kpc,
the surface density and brown dwarf contributions mentioned above, and
a mass $M_d= 2\pi\Sigma_\odot R_d^2 \exp(R_\odot/R_d)\approx 4.6 \times
10^{10}$ $\msol$,

${\bullet} $ a central bar (bulge), whose maximum mass is determined
below, which contributes to the microlensing optical depth with Eq. (25)
of Paper I,

${\bullet}$ a spheroid with a (flattened) De Vaucouleurs profile and a
mass density deduced from the mass function given by Eq. (16) of paper
I. This yields a mass $M_{\rm sph}=1.3\times 10^{10}\,\msol$, which
contributes $\sim 1\%$ of the local dark matter density and $\sim $0.1\%
of the disk total density (Chabrier \& M\'era, 1997a), and an optical
depth $\tau_{\rm sph} \sim 5\times 10^{-9}$ towards the LMC.

${\bullet}$ a screened standard halo described by Eq. (\ref{rhoh}), which
corresponds to a halo mass (see \S 2 of Paper I) $M_h=\vrot^2 r/G\approx
10^{12}(r/100\,$kpc$)\,\msol$. As shown in paper I (see also Chabrier \&
M\'era, 1997a) the contribution of main sequence stars and brown dwarfs
to the halo mass budget is negligible, if not zero.

In this model, one has to invoke a substantial amount of dark matter of
unknown origin.

\begin{figure} \resizebox{\hsize}{!}{\includegraphics{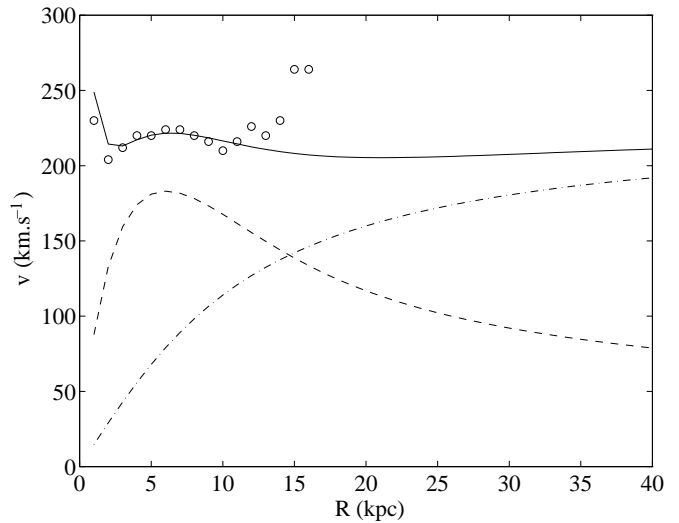}}
\caption{Rotation curve of the standard model (see text). The open
circles correspond to the averaged observed values (from Vall\'ee,
1994), the dashed line is the disk contribution, the dash-doted line the
halo contribution, and the solid line is the total model rotation
curve.\label{figrc1}}
\end{figure}

This model gives the rotation curve shown in figure \ref{figrc1}. We see
immediately that the contribution of the central bar to the velocity
at 1 kpc from the Galactic centre cannot exceed the observed value of
230 $\kms$.  This sets the mass $M_{\rm bulge}\sim 1.2 \times 10^{10}$
$\msol$ within 1 kpc, under the assumption of spherical symmetry. The
presence of a central bar changes this result to $M_{\rm bulge}\sim 2
\times 10^{10}$ $\msol$ (Zhao et al., 1995).\nocite{ZhaoSpergelRich95}

\bigskip

In summary, the standard halo model is consistent with all observed
properties of the Galaxy, at the price of postulating the existence of
a sizeable fraction of non-baryonic matter in the halo (that is needed
anyway at larger scales if the density parameter of the Universe $\Omega$
is larger than $0.1$). To explain the observed microlensing events towards
the LMC, the only (uncomfortable) possibility is to invoke a substantial
amount of white dwarfs in the halo, which contribute to $\sim 40\%$
of the dynamical mass, providing very particular hypothesis on the age
and the initial mass function of the halo, as discussed in the previous
section. Since, as mentioned above, brown dwarfs and low-mass stars seem
to be excluded as a significant halo population, this model implies that
$\sim 60\%$  of the Galactic missing mass is in the form of undetected
{\it non-baryonic} matter. This fraction may be as high as $90\%$ if
the lower limit for the baryonic fraction in the halo inferred from
the number of microlensing events is retained.  Note, however, that
this model does not {\it need} a white dwarf component in the halo.
Would the future number of microlensing events towards the LMC decrease,
or even vanish, this model remains valid.

\section{Maximal disk}

It has been proposed recently on various grounds (e.g. Pfenniger,
Combes \& Martinet, 1994)\nocite{PfennigerCombesMartinet94} that,
contrary to the original suggestion of Ostriker and Peebles
(1973),\nocite{OstrikerPeebles73} the Galaxy has no massive dark halo.
In this case, {\it all} dark matter is distributed in a strongly
flattened, disk-like structure.

It is well known that an exponential disk gives rise to a decreasing
rotation curve, incompatible with the observed flat rotation curve
(see e.g. Binney \& Tremaine, 1987).\nocite{BinneyTremaine87} The
inverse problem, which consists of deriving a density profile from an
observed rotation curve, can be solved analytically for the peculiar
case of a constant rotation velocity (Mestel, 1963).\nocite{Mestel63}
For an {\it infinite} disk, the surface density is proportional to the
radial distance $1/R$:

\beq \Sigma(R) = \Sigma_\odot \frac{R_\odot}{R}\label{SigMestel}\eeq
with

\beq\Sigma_\odot =  208\,  ( \frac{\vrot}{220 \kms} )^2\,\,\mssurf
\label{SigsolMestel}\eeq

This value is much larger than the one inferred from the observed
vertical acceleration (see Paper I).

As mentioned above, Eq. (\ref{SigMestel}) holds for an {\it infinite}
disk. The {\it finite-size} yields a smaller value for $\Sigma_\odot$
for the same rotation curve. In this case, the density profile is
very close to an exponential, $\Sigma(R)\approx \Sigma_0e^{-R/R_d}$,
with a scale length $R_d\approx 1/3\, R_G$, where $R_G$ is the Galactic
(disk) radius (M\'era, Mizony \& Baillon, 1997).\nocite{MeraMizony97}
Only the very central and external parts of such a maximal disk depart
from the exponential profile, as expected. In the centre, this departure
corresponds to the bulge whereas at the edge, the density must decrease
to zero. Thus, for a {\it truncated disk}, a flat rotation curve does not
correspond to a $1/R$ surface density but rather to an {\it exponential}
disk profile plus a central bulge (see M\'era, Mizony \& Baillon, 1997).
The scale length of this exponential is $R_d\sim$ 10 kpc for spiral
galaxies similar to the Milky Way, which corresponds to an exponential
variation of the mass-to-light ratio $M/L$ in the Galactic disk, and an
increasing dark matter/visible matter ratio from the centre to the edge
of the disk. As demonstrated by M\'era et al. (1997b), if all the mass
is distributed in a disk, the local surface density is fixed essentially
by the disk radius $R_G$. For the rotation curve to be flat up to at
least $R_G = 30$ kpc, we get:

\beq \Sigma_\odot  \ge  \, 190\, \mssurf \label{Sigsol190}\eeq

As stressed in \S2, the ``observed'' dynamical densities are derived by
measuring $K(z)$ at some altitude $z$, corrected from the standard halo
contribution to obtain the disk density.  If the halo local density is
reduced, the disk surface density implied by a given vertical acceleration
is different (Eq. (\ref{Kofz})). For the standard halo (\S3), the halo
correction yields :

\beq \Sigma_\odot\,\,  =\,\, \frac{K(\langle z\rangle )}{2\pi G}\,\,
-\,\, 20 \,\mssurf \frac{\langle z\rangle}{1\, \rm kpc} \eeq

\noindent where $\langle z\rangle$ is the average height of the sample.

In case there is no halo (maximal disk),  $20\, \mssurf
\frac{\langle z\rangle}{1 \,\rm kpc}$ must be added to the
$\Sigma_\odot$ values quoted by the various authors. With our
estimate of $\langle z\rangle = 1$ kpc, 1.1 kpc, 0.5 kpc and 0.5
kpc for the Bienaym\'e et al (1987),\nocite{BienaymeRobinCreze87}
Kuijken and Guilmore (1991),\nocite{KuijkenGilmore91} Bahcall
et al (1992)\nocite{BahcallFlynnGould92} and Flynn and Fuchs
(1994)\nocite{FlynnFuchs94} surveys,
respectively, the observed values
become $ \Sigma_\odot  = 70 \pm 10 $, $70 \pm 9 $,
$95 \pm 25  $,  $62 \pm 13 \,\mssurf $, respectively, with an
average $ 70 \pm 6 \,\mssurf$
\footnote{As mentioned in Paper I, the Bahcall et al. (1992) and Flynn \& Fuchs (1994) observations are not completely independent but the weight of the Bahcall et al. value is small and droping it does not affect the result.}.
Therefore, the surface density
(\ref{Sigsol190}) needed to comply with the rotation curve is many
standard deviations off the observed value.

One may argue that the vertical distribution of the gas increases
with Galactocentric distances, as expected if the gas is supported by
rotation (Pfenniger et al, 1994). This would yield a dark ``corona''
beyond the edge of the observed optical disk. However, this is not a
satisfactory solution, since, as can be shown easily, adding a dark
corona of matter {\it outside} the disk will {\it lower} the rotation
velocity inside the Galactocentric radius of this corona, so that the
disk surface density must be increased accordingly in order to reproduce
the observed rotation velocity, to a value close to (\ref{SigsolMestel}).

In sum, the total mass of the disk in this model is $M_d\approx
10^{11}\times R_G/10$ kpc. It seems unrealistic that the disk extends
beyond $40$ kpc for stability reasons so that such a description of our
Galaxy would not be consistent with the velocity fields observed up to
$60$ kpc (see \S2 of Paper I). It similarly implies that the mass of the
Milky Way is significantly smaller than the required $2.4 \pm 0.8\times
10^{12}\,\msol$ deduced from the dynamics of the Local Group (Peebles,
1994; see \S2 of Paper I).\nocite{Peebles94} Or it would impose that
almost all the mass of the Local Group ($\sim 4\times 10^{12}$ $\msol$)
is in Andromeda. The dynamics of our Galaxy would be explained without
the presence of non-baryonic dark matter at the price of an unacceptably
large amount of such material in Andromeda.

The only way out is to require {\it the dark matter to be distributed
over a scale height $z>3$ kpc}. In this case, the maximal disk is
similar to a highly flattened halo, a solution apparently not supported
by observations, as mentioned in \S2.  If the sought baryonic matter
is in the form of dark, molecular hydrogen clouds (Pfenniger et al,
1994; Gerhard \& Silk, 1996), this raises a severe problem, for the
observed distribution of {\it atomic} hydrogen (Dickey and Lockman,
1990)\nocite{DickeyLockman90} is well confined within 200 pc from the
Galactic plane. This implies that molecular and atomic hydrogen gas must
have a completely different (a factor 10) Galactic scale height.

\medskip

In summary, the maximal disk or highly flattened halo models seem
to be excluded by kinematic observations and a 3-dimensional dark
halo contribution is necessary, which is responsible for the constant
(non-keplerian) rotation curve beyond the disk truncation radius.

\section{Modified halo model}

As shown in Eq. (\ref{rhoh}), the halo distribution is determined by
two parameters: its size, fixed by the total Galactic mass, and its core
radius $R_c$.  This latter is not physically motivated but is determined
by the condition that the visible plus dark matter distributions
give rise to the measured rotation curve, which yields the local halo
normalization. The standard halo model (\S 3) assumes by definition
that the dark matter is mostly in the dark halo, so that the local halo
density is $\rho_{h_\odot}=9\times 10^{-3}\,\msvol$ and $R_c= 5\pm 1$ kpc
(Eq. \ref{rhohsol}).

We modify the standard model by putting as much baryonic mass in the
disk as allowed by the microlensing observations and the star counts,
from the analysis conducted in Paper I.  The aim is to derive the maximum
limit of a predominantly baryonic model for the Galaxy compatible with all
presently available observations.

We first note that there is observational evidence that
the optical disk is truncated at a galactocentric distance
$R\approx 14$ kpc (Robin, Cr\'ez\'e \& Mohan, 1992; Ruphy et al.,
1996).\nocite{RobinCrezeMohan92,Ruphyetal96} As mentioned in
section \S4, such a truncated disk plus a central bulge yield a flat rotation
curve up to the limit of the disk, with a scale length $R_d\approx 3.3$
kpc, in agreement with observations ($1.8< R_d < 5$ kpc, see Kent, Dame \&
Fazio, 1991)\nocite{KentDameFazio91,MeraMizony97} and $\Sigma_\odot\approx
120\, \mssurf$ (M\'era et al., 1997b). Since a maximal disk is excluded
(\S4), we must include a halo contribution given by the following
flattened distribution (in cylindrical coordinates):

\beq \rho_h (R,z)=\rho_{h_\odot} {R_\odot^2+R_c^2\over R^2+z^2/q^2+R_c^2}
\label{rhohf}\eeq

\noindent with

\beq \rho_{h_\odot} = f\times \frac{\vrot^2}{4\pi G}
\frac{1}{R_c^2+R_\odot^2} \frac{\sqrt{1-q^2}}{q\, {\rm Arccos} q},\eeq

\noindent where $f$ is the normalization of the halo, and $q\sim 0.6$.
Straightforward calculations yield the following expression for the
contribution of such a halo to the rotation curve:

\beq v_h^2=f\,\vrot^2\left[1 -
\frac{R_c\,\epsilon}{\sqrt{R_c^2\,\epsilon^2+R^2}} \frac{\mbox{Arctan}
\frac{1}{q}\sqrt{\epsilon^2+R^2/R_c^2}}{\mbox{Arcsin}\,\epsilon}
\right]\label{vhalo}\eeq

\noindent where $\epsilon=\sqrt{1-q^2}$. For $R>>R_c$, the asymptotic
rotation velocity is $\sqrt{f}\vrot$, with $\vrot \approx 220\,\kms$.

The core radius $R_c$  plays in fact the same role as the normalization
factor $f$, as shown by eqn. (\ref{rhohf}). Halo models ($f=0.35; R_c=5$
kpc) and ($f=1; R_c=15$ kpc) yield the same normalization in the solar
neighborhood $\rho_{h_\odot}=3\times 10^{-3}\,\msvol$, i.e. $\sim$
30\% of the standard model local density, but the second one yields the
{\it correct rotation velocity at large distances} ($R\wig > 20$ kpc).
A truncated disk plus a dark halo with a large core-radius, $R_c\sim 15$
kpc is thus a solution worth considering.  This model is consistent
with the microlensing experiments towards the LMC which yield a dark
matter contribution $\sim 40\%$ of the standard halo: the microlensing
optical depth toward the LMC derived from the afore-mentioned density
is $\tau=3\times 10^{-7}$, in agreement with the observed value.

\begin{figure}\resizebox{\hsize}{!}{\includegraphics{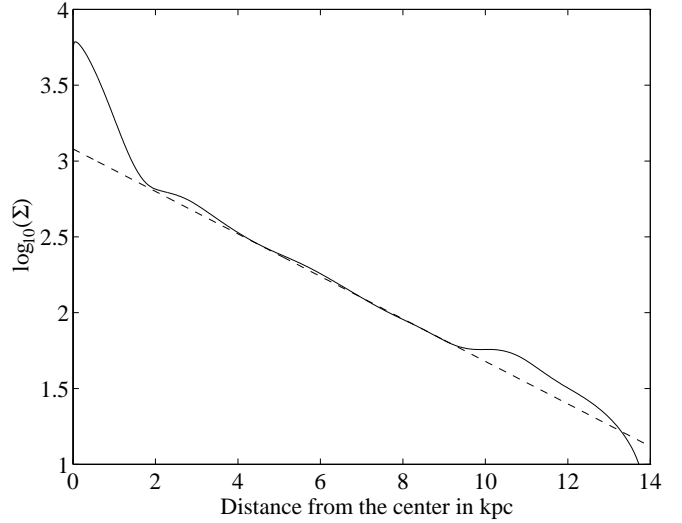}} 
\caption{Surface density profile of the Galactic disk inferred from the
observed rotation curve (Vall\'ee, 1994), from which the modified-halo
($R_c=15$ kpc) contribution has been removed (see text). The dashed line
is an exponential profile of scale length 3.2 kpc, and solar density $87\
\mssurf$.} \end{figure}

We have substracted the contribution of this modified halo from the
observed rotation curve, taken from Vall\'ee (1994), up to $\sim 14$
kpc. Figure 2 shows the inferred density profile for the disk. The
agreement between theory and observation yields a scale length $R_d=3.2$
kpc and a disk mean surface density $\Sigma_\odot=87\,\mssurf$.

\begin{figure} \resizebox{\hsize}{!}{\includegraphics{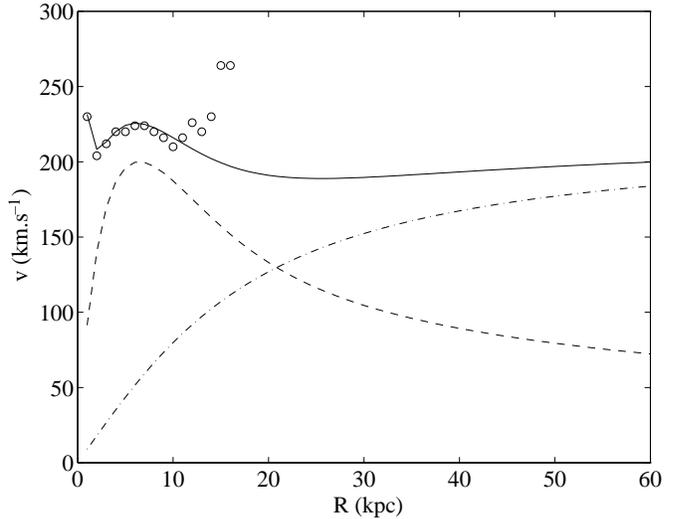}}
\caption{Rotation curve of the modified halo model. The disk
contribution corresponds to the dashed line, while the halo contribution
corresponds to the dash-dotted line.
The solid line is the total
rotation curve, with the observed rotation curve (Vall\'ee, 1994) for $R<14$
kpc. Note that the observational uncertainties become very large beyond 10 kpc (see Fich \& Tremaine, 1991).} \end{figure}

Instead of using the numerical density profile derived from the inversion technique of M\'era et al. (1997b), we will use the more simple exponential density law. Then the observed rotation curve is well reproduced with the following disk density:

$$ \Sigma_d (R) = (75\ \mssurf)\times \ e^{R_\odot/R_d} \ e^{-R/R_d}$$

Figure 3 shows the total (disk + halo) rotation curve up to 60 kpc. The
disk dominates the rotation curve up to its edge, i.e. $\sim 14$ kpc,
whereas the dark halo becomes dominant beyond this limit and reaches
the correct asymptotic value 220 $\kms$. The rotation curve of the
model departs significantly from the observations for $R>12$ kpc,
but the uncertainty begins to be large at these radii (about 20\%). In
particular, the last two points are derived from 5 objects only, each
with large errors. Therefore, we can consider the model as consistent
with the observed rotation curve.

The most interesting property of this model is that it is consistent
with a {\it fully baryonic} dark matter solution, most of the matter
being in the dark halo, and some of it in the disk.  The disk dark
population consists of brown dwarfs (see \S 4.1.2 of Paper I), whereas
for the halo, the most likely candidates are white
dwarfs, as discussed in \S2. As discussed at length in \S3.1 and \S4.1
of Paper I, a substantial population of disk brown dwarfs is consistent with
present-day microlensing observations toward the Galactic center. It is
also consistent with the extrapolation into the sub-stellar domain of the
M-dwarf mass function obtained either from the nearby luminosity function
(Kroupa, 1995) or from the very faint end of the HST luminosity function
(Gould et al., 1997), if the apparent upturn in this region is real.
Such mass functions $dN/dm\propto m^{-\alpha}$ have a slope $\alpha
\wig > 2$ and a normalization near the bottom of the main sequence
$dN/dm(0.1\,\msol)\sim 1\,\msol^{-1}pc^{-3}$ (see \S3.1 of Paper I).

A rising MF near the brown dwarf domain is supported by the recent study
of the distribution of secondary masses of 20 F and G dwarf binaries by
Mazeh, Latham \& Stefanik (1996), although poor statistics (3 detections)
preclude the derivation of a precise MF.  On the other hand, a shallower
MF or a substantially smaller normalization near the bottom of the MS
will rule out the possibility of a substantial brown dwarf mass-fraction
in the Galactic disk. Therefore nailing down the issue on the very end
of the disk MF bears essential consequences for Galactic modelling. This
should be done shortly with the DENIS whole-sky survey.

As shown above, the main problem of such a modified-halo model is the
large inferred surface density $\Sigma_\odot\sim 75\,\mssurf$. For this model,
the halo correction is only

\beq \Sigma_\odot \,\, = \,\,  \frac{K(\langle z\rangle)}{2\pi G} \,\, -
\,\, 6\, \mssurf \frac{\langle z\rangle}{1\,
{\rm kpc}},\eeq

\noindent so we have to increase by $ 14$ (20 minus 6) $\mssurf
\frac{\langle z\rangle}{1\ \mbox{kpc}} $ the observed dynamical  values. This
yields $ \Sigma_\odot  = 63 \pm 10,\ 62 \pm 9,\ 91
\pm 25,\ 58 \pm 13\ \mssurf $ for Bienaym\'e et al (1987),
Kuijken and Gilmore (1989), Bahcall et al (1992), Flynn and Fuchs
(1994) determinations, respectively, with the average $ \Sigma_\odot
= 63 \pm 6\, \mssurf$.  Since the detected surface density is
$\Sigma_{\odot_{vis}}=43\pm 5\, \mssurf$ (Eq. 15 of Paper I), the brown
dwarf surface density in the solar neighborhood is then $\Sigma_{\rm bd}
= 15\pm 14$, still compatible with microlensing observations (see \S4.1.2
of Paper I).

The axisymmetric mean density at the solar radius $\Sigma_\odot
\approx 75\, \mssurf$ obtained in this model is larger by $\sim 1.3$
standard deviations than the afore-mentioned Flynn \& Fuchs (1994)
recent determination in the solar neighborhood. The difference, however,
may be attributed to the fact that the Sun is located in-between two
spiral arms, namely $\sim 1.5$ kpc from the Sagittarius arm and $\sim
2$ kpc from the Perseus arm (King, 1989).\nocite{King89} Therefore
the solar surface density might be slightly smaller than the averaged
surface density along the circle of radius $R_\odot$. This is supported
by the observations of external galaxies and by numerical simulations,
which give about a factor 2 for the maximal/minimal density in the disk,
expected to translate into a smaller factor for the vertical acceleration
$K(z)$ (see e.g. Rix and Zaritsky, 1995). The value at 1.1 kpc of $K_z/2\pi G$ derived by Kuijken and Gilmore (1991) is $71\pm 6 \ \mssurf$, whereas for the present heavy disk model $K_z/2\pi G = 81\ \mssurf$, corresponding to a ratio of 1.14, which could be accounted for by the spiral structure.

The average density we need
is $\sim$ 1.3 times larger than the mean solar value, and only 5\% times
larger than the 1-$\sigma$ upper limit of the Flynn and Fuchs (1994)
value. Although the relevance of such values deserves a detailed study,
they do not seem to be unrealistic.

\section{Summary and conclusions}

In the present paper, we have derived consistent mass-models for
the Galaxy, disk(s), bulge, spheroid and dark halo. The present
calculations differ substantially from previous approaches
which relied on microlensing analysis only (e.g. Alcock et al.,
1996b)\nocite{Alcocketal96b} or which were based on the minimization
of models with a large number of free parameters (e.g. Gates, Gyuk and
Turner, 1996)\nocite{GatesGyukTurner96}. We proceed differently by first
determining as accurately as possible the main parameters entering the
Galactic modelling from {\it all} observational constraints presently
available, i.e. star counts and luminosity functions of metal-rich and
metal-poor M-dwarfs at faint magnitudes, including the most recent Hubble
Deep Field data, microlensing observations towards the LMC and the central
part of the Galaxy, including not only the optical depths but the much
more constraining time-distributions of the events, and large-scale
(circular rotation velocity) and small-scale (local gravitational
acceleration) kinematic properties. The microlensing analysis is
derived not only in term of optical depth but includes the complete
time-distributions of the events.  These calculations include previous
results wich yield the determination of the mass-functions in the Galactic
disk (M\'era, Chabrier \& Baraffe, 1996)\nocite{MeraChabrierBaraffe96}
and spheroid (M\'era, Chabrier \& Schaeffer, 1996; Chabrier \& M\'era,
1997a)\nocite{MeraChabrierSchaeffer96,ChabrierMera97a} and of the
contribution of stellar and substellar objects to the Galactic disk and
halo mass budgets (M\'era, Chabrier \& Schaeffer, 1997; Paper I; Chabrier
\& M\'era, 1997a).\nocite{Meraetal97a} This yields a {\it consistent}
determination of the different parameters.

We show that a maximal disk model is clearly excluded by the observations,
in particular by the observed local surface density. An extension of this
model, where dark matter is in the form of giant molecular clouds located
at the very periphery of the Galactic disk (Pfenniger et al, 1994) raises
severe problems : i) it implies that molecular hydrogen is distributed
over a scale {\it height} about 10 times larger than the observed {\it
atomic} hydrogen distribution, for the surface density to agree with the
observed value, ii) it implies a scale {\it length} for the disk $R_d\sim
10$ kpc, about 3 times larger than the value deduced from star counts,
for the circular rotation velocity to be constant up to $\sim 30$ kpc,
iii) it does not explain the observed microlensing events toward the LMC.
In case such clouds are distributed in the halo (Gerhard \& Silk, 1996),
point i) and iii) remain valid. An extension of this model, namely a
highly flattened ($q\sim 0.1$) halo, although not completely ruled out,
is at odd with various observational constraints (see \S 2).

Two models (and possibly a whole range of models in-between these
extremes) fulfill {\it all} the afore-mentioned observational
contraints. The various components of these two models are given in
Table I. Both include:

\indent$\bullet $ a central, elongated bulge, with $M_{\rm bulge}\sim
2\times 10^{10}\,\msol$ from kinematic considerations. The analysis of
the first year of MACHO observations towards the bulge (Alcock et al.,
1997)\nocite{Alcocketal97} seems to exclude a significant population
of brown dwarfs in this region (Paper I) although better statistics are
needed to really nail down this issue,

\indent$\bullet $ a thin double-exponential disk with a scale
length $R_1\sim 3$ kpc and a scale height $h_1\sim 320$ pc (Bahcall
\& Soneira, 1980)\nocite{BahcallSoneira80} (or equivalently a ${\rm
sech}^2\times\exp$ model (Gould et~al., 1997)), + a thick ${\rm
sech}^2\,(|z|/h_2)$ disk with the same scale length, a scale height
$h_2\sim$ 640 kpc and a local normalization $\sim 20\%$ (Gould et~al.,
1997). The mass of the disk reads:

\beq M_{\mbox{disk}}\approx 2\pi \Sigma_\odot R_d^2
e^{R_\odot/R_d}\label{Mdisk}\eeq

\noindent where $\Sigma_\odot$ is the local mass density, which is
different in the two models.

\indent$\bullet $ a flattened ($q\sim 0.6$)  DeVaucouleurs spheroid with
a mass $M_{\rm sph}\approx 1.3\times 10^{10}\,\msol$, i.e. a negligible
contribution to the Galactic mass budget (Chabrier \& M\'era, 1997a),

\indent$\bullet $ a flattened dark halo with the density profile
(\ref{rhohf}), which yields a mass:

\begin{eqnarray} M_h & = & \frac{\vrot^2}{G}(R_h-R_c\,\mbox{Arctan}
{R_h\over R_c})  \nonumber \\
 & \approx  &
1.36\times 10^{12}\,{R_h\,+\,23\,{\rm kpc} \over 100\,+\, 23\,{\rm
kpc}}\,\,\msol, \label{Mhalo} \end{eqnarray}

\noindent which is nearly independent of the value of the core radius
$R_c$.

In both models, a significant contribution of low-mass stars and brown
dwarfs to the dark halo mass is excluded from star count observations and
microlensing experiments (Chabrier \& M\'era, 1997a; M\'era, Chabrier
\& Schaeffer, 1997). Although we can argue that neither the EROS nor
the MACHO experiments are optimized for short-duration ($\sim$ days)
observations\footnote{Although the EROS CCD experiment was optimized for short timescales, the exposure, $\sim 80000$ stars $\times$ a few months, is small.}, a significant population of low-mass (brown dwarfs) objects
in the halo would produce {\it many} short-time events: $N=\Gamma \times
E\times \epsilon$, where $E=1.8\times 10^7$ star.year is the exposure of
the MACHO collaboration, $\epsilon$ is their mean detection efficiency and
$\Gamma \sim 1.6\times 10^{-6}\,(m/\msol)^{-1/2}$ event/year is the event
rate for the standard halo model. A Dirac MF in the BD domain $m\wig <
0.1\msol$ implies about 16 events with $t_e\wig <$ 20 days, which is
excluded at 99\% confidence level, since only one has been observed (with
$t_e\approx 19.4$ days, Alcock et al., 1996b). Planetary-like objects ($m<
10^{-2}\,\msol$) are also excluded as a significant mass-contribution
($\wig > 10\%$) at 95\% confidence level by the EROS CCD observations
(Renault et al., 1997)\nocite{Renaultetal97} and the MACHO ``spike''
observations (Alcock et al., 1996a).\nocite{Alcocketal96a} In both models
also, the dark halo contains a significant population of compact objects,
which are responsible for the microlensing observations towards the LMC
and amount to a local density $\rho \sim 3\times 10^{-3}\,\msvol$. As
mentioned below, this condition is compulsory for the second
(modified halo) model. Although uncomfortable, the halo WD solution
represents the ``least unlikely'' scenario and cannot be excluded at
the time these lines are written.

\bigskip
\begin{table}
\caption[]{Main characteristic of the two consistent Galactic models
outlined in the Conclusion.
The dark halo is made of white dwarfs in both models, but the standard model
has a dominant non baryonic component.}
\begin{tabular}{lccl}
\hline
\noalign{\smallskip}
\mbox{} & Std halo & heavy disk & unit \\
\noalign{\smallskip}
\hline
\noalign{\smallskip}
Disk: & & & \\
\mbox{}\hspace{0.5cm} Scale length & 3.2 & 3.2 & kpc\\
\mbox{}\hspace{0.5cm} Scale height & 320 & 320 & pc\\
\mbox{}\hspace{0.5cm} $\rho_{\rm bd}$ & $\wig < 0.015$ & $\sim 0.03$ &
$\msvol$ \\
\mbox{}\hspace{0.5cm} Mass & $3\times 10^{10}$ & $5\times 10^{10}$ &
$\msol$\\
Thick Disk: & & &\\
\mbox{}\hspace{0.5cm} Scale length & 3.2 & 3.2 & kpc\\
\mbox{}\hspace{0.5cm} Scale height & 640 & 640 & pc\\
\mbox{}\hspace{0.5cm} Mass & $ 6\times 10^9$ & $ 6\times 10^9$ & $\msol$
\\
Bulge: & & & \\
\mbox{}\hspace{0.5cm} Mass & $1.2 \times10^{10}$ & $ 1.2\times10^{10}$ &
$\msol$ \\
\mbox{}\hspace{0.5cm} Bar & yes & yes & \\
\mbox{}\hspace{0.5cm} Brown dwarfs & none & none &  \\
Spheroid: & & & \\
\mbox{}\hspace{0.5cm} Density & De Vaucouleurs & De Vaucouleurs & \\
\mbox{}\hspace{0.5cm} Oblateness & $q\sim 0.6$ & $q\sim 0.6$ & \\
\mbox{}\hspace{0.5cm} Mass & $1.3\times 10^{10}$ & $1.3\times 10^{10}$ &
$\msol$ \\
Halo: & & & \\
\mbox{}\hspace{0.5cm} Core radius & 5 & 14 & kpc \\
\mbox{}\hspace{0.5cm} Composition & $<50$\% WD & up to 100\% WD & \\
\mbox{}\hspace{0.5cm} Mass & $\sim 10^{12}$ & $\sim 10^{12}$ & $\msol$
\\
\mbox{}\hspace{0.5cm} $v_\infty$ & 220 & 220 & $\kms$ \\
Total column density: & & & \\
$\langle \Sigma \rangle$ at $R_\odot$ & $\sim 55$ & $\sim 75$ &
$\mssurf$ \\
\noalign{\smallskip}
\hline
\end{tabular}
\end{table}
\bigskip

The very difference between these two models resides in the location of
the Galactic dark matter. The first model is simply what we call the
``standard'' model with a {\it possible} ($\sim 10\pm 10 \,\mssurf$)
amount of hidden mass under the form of brown dwarfs in the disk or in
the bulge. In this case, {\it nearly all} dark matter is in the dark
halo and is predominantly ($> 50\%$) of {\it non-baryonic nature}. In the
second case, the Galactic disk must necessarily include a significant
brown dwarf population.  The analysis conducted in Paper I (\S4.1.2)
yields a brown dwarf contribution $\Sigma_{\odot_{\rm BD}} \sim 20-30
\mssurf$, comparable to the stellar contribution, and thus a local surface
density $\Sigma_\odot \simeq 60-70 \,\mssurf$, since $\Sigma_{\odot_{\rm
vis}}=43\,\pm\,5\, \mssurf$ (Eq. (15) of Paper I), within the 1$\sigma$
limits of the observed dynamical value (Flynn \& Fuchs, 1994). The
{\it average} disk surface density at the solar radius is in this model
$\sim 5-20\%$ larger than the dynamical value determined in the solar
neighborhood. This may be due to the presence of density waves.
Most of the dark matter is still in the dark halo but its contribution to the
local dynamical density amounts only to $\sim 3\times 10^{-3}\,\msvol$,
$\sim 30\%$ of the {\it standard} halo contribution. This (heavy disk)
model necessarily needs the presence of white dwarfs (or other remnants)
in the halo to be confirmed, these latter providing most of the missing
mass. In this case, the missing mass in spiral galaxies is {\it entirely
baryonic} and the dark baryons consist essentially of brown dwarfs
in the disks and white dwarfs in the halos. This model, however, must
be considered as an extreme case. But it shows the possibility of a
heavy Galactic disk, with a substantial fraction of brown dwarfs, which yields
a smaller local normalization for the halo than assumed in the usual
standard disk+halo model. This leads to a {\it predominantly}, rather than
``entirely" baryonic model, a more likely solution.  This model, however,
requires a rather large surface density in the solar neighborhood and a
relatively steep ($\alpha \sim 2$) mass function in the disk near the
hydrogen-burning limit, which extends apparently (within the present
uncertainties) to smaller masses than in the bulge (see Paper I, \S
4.1.1 and 4.1.2).

Therefore, the key issue to determine the correct Galactic mass-model,
based on the present analysis, is the determination of i) the brown dwarf
density in the disk/bulge and ii) the white dwarf (or other candidates)
density in the dark halo. Ongoing microlensing experiments (EROS~II,
MACHO), infrared surveys and large field high-proper motion surveys
might soon nail down this issue. Analysis of these observations in the
context of the present models will yield immediatly the contributions of
the various baryonic components to the different parts of the Galaxy and
eventually the characterization of the (at least baryonic) dark matter
in the Galaxy.  Meanwhile, the DENIS and 2MASS projects should allow a
precise determination of the disk stellar mass function down to (perhaps
below) the hydrogen-burning limit, allowing more robust extrapolation into
the brown dwarf domain.  Unfortunately, observations at the time these
lines are written do not allow to disentangle between these two models.
Various uncertainties in present-day observations (scale heights, scale
lengths, $\vrot, \Sigma_\odot, R_\odot$, microlensing observations)
remain within the error bars allowed by the present models.

We have shown in the present paper that both types of models yield the
observed, constant circular rotation curve up to large galactocentric
distances. In particular a truncated disk yields an {\it exponential}
density profile. In the second model, the disk contributes to the
rotation curve up to its edge, and the dark halo beyond this limit. Both
the disk and the halo yield a flat rotation curve and there is {\it no
need anymore to invoke any peculiar conspiracy}.  We stress that, in
virtue of the Occam's razor principle, we have considered in the present
calculations the most natural hypothesis for the nature and the location
of dark matter.  We have assumed in particular that the dark matter
is distributed homogeneously, as visible matter.  This yields about
constant, but different, mass-to-light ratios in the disk and in the
halo. We have not considered more exotic scenarios involving clustered
dark matter in the halo (see e.g. Kerins, 1997).\nocite{Kerins97} We
believe the present calculations to provide new insight on the possible
nature and distribution of dark matter in galaxies. Comparisons of
future observations with the present models should help answering this
long-standing problem.

\bigskip 

{\bf Acknowledgments:} We have benefited all  along this work of
numerous discussions with various people, more particularly with the
MACHO and the EROS groups and with D. Pfenniger and F. Combes. We also
acknowledge useful conversations with O. Gerhard and with the referee,
A. Gould. It is a pleasure to thank these persons for their valuable
help.

\bigskip 

{\it Note added in proof}. After the present calculations were completed, a new determination of the local dynamical density has been derived from the Hipparcos data (Cr\'ez\'e et al., astro-ph/9709022). This new determination leaves essentially no room for any dark component in the disk and thus, if confirmed, seems to exclude the modified-halo model (\S 5) and to confirm the standard-halo model (\S 3).

\bibliographystyle{astron}

\end{document}